\documentclass[aps,pre,twocolumn,superscriptaddress,showpacs]{revtex4}
\usepackage{epsfig,graphicx}
\usepackage{dcolumn}
\usepackage{bm}
\usepackage{amsmath}
\usepackage{times}

\begin{document}

\title{Emergence of grouping in multi-resource minority game dynamics}

\author{Zi-Gang Huang}
\affiliation{Institute of Computational Physics and Complex Systems,
Lanzhou University, Lanzhou Gansu 730000, China}

\author{Ji-Qiang Zhang}
\affiliation{Institute of Computational Physics and Complex Systems,
Lanzhou University, Lanzhou Gansu 730000, China}

\author{Jia-Qi Dong}
\affiliation{Institute of Computational Physics and Complex Systems,
Lanzhou University, Lanzhou Gansu 730000, China}

\author{Liang Huang}\email{huangl@lzu.edu.cn}
\affiliation{Institute of Computational Physics and Complex Systems,
Lanzhou University, Lanzhou Gansu 730000, China}
\affiliation{School of Electrical, Computer, and Energy
Engineering, Arizona State University, Tempe, AZ 85287, USA}

\author{Ying-Cheng Lai}
\affiliation{School of Electrical, Computer, and Energy
Engineering, Arizona State University, Tempe, AZ 85287, USA}
\affiliation{Department of Physics, Arizona State University,
Tempe, AZ 85287, USA}

\date\today

\begin{abstract}

The Minority Game (MG) has become a paradigm to probe complex
social and economical phenomena where adaptive agents compete for
a limited resource, and it finds applications in statistical and
nonlinear physics as well. In the traditional MG model, agents are
assumed to have access to global information about the past
history of the underlying system, and they react by choosing one
of the two available options associated with a single resource.
Complex systems arising in a modern society, however, can possess
many resources so that the number of available
strategies/resources can be multiple. We propose a class of models
to investigate MG dynamics with multiple strategies. In
particular, in such a system, at any time an agent can either
choose a minority strategy (say with probability $p$) based on
available local information or simply choose a strategy randomly
(with probability $1-p$). The parameter $p$ thus defines the {\em
minority-preference probability}, which is key to the dynamics of
the underlying system. A striking finding is the emergence of
strategy-grouping states where a particular number of agents
choose a particular subset of strategies. We develop an analytic
theory based on the mean-field framework to understand the
``bifurcation'' to the grouping states and their evolution. The
grouping phenomenon has also been revealed in a real-world example
of the subsystem of $27$ stocks in the Shanghai Stock Market's
Steel Plate. Our work demonstrates that complex systems following
the MG rules can spontaneously self-organize themselves into
certain divided states, and our model represents a basic
mathematical framework to address this kind of phenomena in
social, economical, and even political systems.

\end{abstract}

\pacs{02.50.Le, 89.75.Hc, 87.23.Ge}

\maketitle

\section{Introduction} \label{sec:intro}

The Minority Game (MG) was originated from the El Farol bar
problem in game theory first conceived by Arthur in 1994
\cite{Arthur:1994}, where a finite population of people try to
decide, at the same time, whether to go to the bar on a particular
night. Since the capacity of the bar is limited, it can only
accommodate a small fraction of all who are interested. If many
people choose to go to the bar, it will be crowded, depriving the
people of the fun and thereby defying the purpose of going to the
bar. In this case, those who choose to stay home are the winners.
However, if many people decide to stay at home then the bar will
be empty, so those who choose to go to the bar will have fun and
they are the winners. Apparently, no matter what method each
person uses to make a decision, the option taken by majority of
people is guaranteed to fail and the winners are those that choose
the minority one. Indeed, it can be proved that, for the El Farol
bar problem there are mixed strategies and a Nash-equilibrium
solution does exist, in which the option taken by minority wins
\cite{Gintis:2009}. A variant of the problem was subsequently
proposed by Challet and Zhang, named as an MG problem
\cite{CZ:1997}, where a player among an odd number of players
chooses one of the two options at each time step. Subsequently,
the model was studied in a series of works
\cite{CM:1999,CMZ:2000,MMM:2004,BMM:2007,RMR:1999,PBC:2000,ZWZYL:2005,EZ:2000,KSB:2000,Slanina:2000,ATBK:2004,JHH:1999,HJJH:2001,LCHJ:2004,TCHJ:2005,CMM:2008,BMFM:2008,XWHZ:2005,ZZZH:2005}.
In physics, MG has received a great deal of attention from the
statistical-mechanics community, especially in terms of problems
associated with non-equilibrium phase transitions
\cite{Moro:2004,CMZ:2005,YZ:2008}.

In the current literature, the setting of MG is that there is a
{\em single} resource with players' two possible options (e.g., in
the El Farol bar problem there is a single bar and the options of
agents are either going to the bar or not), and an agent is
assumed to react to available global information about the history
of the system by taking on an alternative option that is different
than the current one it is taking. The outstanding question
remains of the nonlinear dynamics of MG with {\em multiple
resources}. The purpose of this paper is to present a class of
multi-resource MG models. In particular, we assume a complex
system with multiple resources and, at any time, an individual
agent has $k > 1$ \emph{resources/strategies} to choose from. We
introduce a parameter $p$, which is the probability that each
agent responds based on its available local information by
selecting a less crowded resource in an attempt to gain
higher payoff. We call $p$ the {\em minority-preference
probability}. We find, strikingly, as $p$ is increased, the
phenomenon of grouping emerges, where the resources can be
distinctly divided into two groups according to the number of
their attendees. In addition, the number of stable pairs of groups
also increases. We shall show that the grouping phenomenon plays a
fundamental role in shaping the fluctuations of the system. The
phenomenon will be demonstrated numerically and explained by a
comprehensive analytic theory. An application to the analysis of
empirical data from a real financial market will also be
illustrated, where grouping of stocks (resources) appears. Our
model is not only directly relevant to nonlinear and complex
dynamical systems, but also applicable to social and economical
systems.

Our multi-resource MG model is presented in Sec. \ref{sec:model}.
The emergence of grouping phenomenon is demonstrated in Sec.
\ref{sec:numerics}. A general theory is developed in Sec.
\ref{sec:theory} to elucidate the dynamics of the emergence and
evolution of the strategy groups. Concluding remarks are presented
in Sec. \ref{sec:conclusion}.

\section{Multi-resource minority game model} \label{sec:model}

We consider a complex, evolutionary-game type of dynamical system
of $N$ interacting agents competing for multiple resources. Each
agent will chose one resource in each round of the game. And, each
resource has a limited capacity, i.e., the number of agents it can
accommodate has an upper bound $n_c$. There are thus multiple
strategies ($s=1$, $2$, $\cdots$, $k$, where $k$ is the maximum
number of resources/strategies) available to each agent. On
average, each strategy can accommodate $N/k$ agents, and we
consider the simple case of $n_c=N/k$. Let $n_s$ be the number of
agents selecting a particular strategy $s$. If $n_s \le n_c$, the
corresponding agents win the game and, consequently, $s$ is the
{\em minority strategy}. However, if $n_s>n_c$, the associated
resource is too crowded so that the strategy fails and the agents
taking it lose the game, which defines the ``majority strategy.''
The optimal solution to the game dynamics is thus $n_s=n_c$.

In a real-world system, it is often difficult or even impossible
for each agent to gain global information about the dynamical
state of the whole system. It is therefore useful to introduce the
concept of \emph{local information network} in our
multiple-resource MG model. At each time step, with probability
$p$, namely the \emph{minority-preference probability}, each agent
acts based on local information that it gains by selecting one of
the $k$ available strategies. In contrast, with probability $1-p$,
an agent acts without the guidance of any local information. For
the minority-preference case, agent $i$ has $d$ neighbors in the
networked system. The required information for $i$ to react
consists of all its neighbors' strategies and, among them, the
winners of the game, i.e., those neighboring agents choosing the
minority strategies at the last time step. Let $\Pi=\{s_m \}$ be
the set of minority strategies for $i$'s winning neighbors, where
a strategy may appear a number of times, if it has been chosen by
different winning neighbors. With probability $p$, agent $i$ will
chose one strategy randomly from $\Pi$. Thus, the probability
$P_s$ for strategy $s$ to be selected is proportional to the times
it appears in $\Pi$, i.e., $P_s=N_{s}/Card(\Pi)$, where
$Card(\Pi)$ is the number of elements in $\Pi$ and $N_{s}$ the
times strategy $s$ appears in $\Pi$. If $\Pi$ is empty, $i$ will
randomly select one from the $k$ available strategies. While, for
the case that an agent selects a strategy without the guidance of
any local information with probability $1-p$, it will either
choose a different strategy randomly from the $k$ available ones
with mutation probability $m$ \cite{ZWZYL:2005}, or inherit its
strategy from the last time step with probability $1-m$.

\section{Numerical results} \label{sec:numerics}

\begin{figure*}
\epsfig{figure=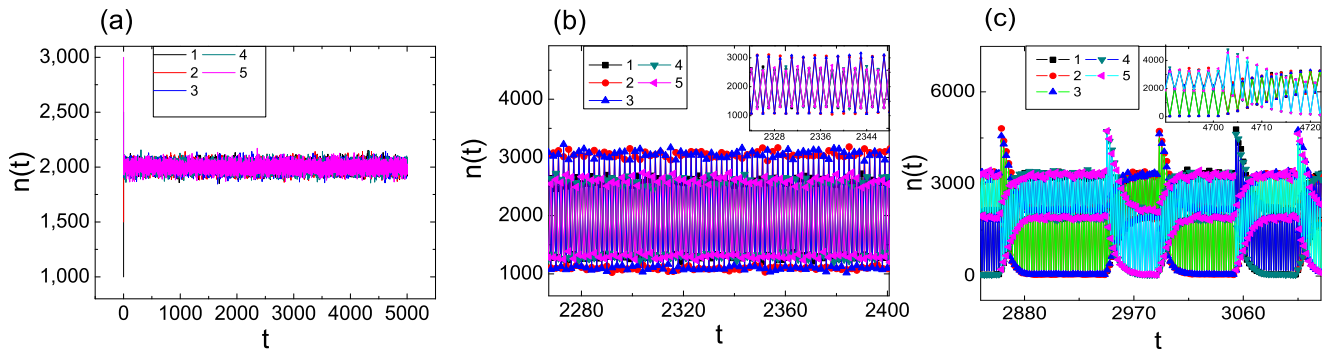,width=\linewidth}
\caption{(Color online.) (a-b) For a square-lattice system of
$N = 10000$ agents, time series of the number of agents selecting each
of the $k = 5$ available strategies for $p=0$, 0.45, and 1.0, respectively.
The probability of random alteration of strategy in the case of
complete lack of local information is set to be $m=1$. A strategy
grouping behavior can be seen in (b), where the whole strategy set
is broken into two groups: one of two and another of three strategies.}
\label{fig:ts}
\end{figure*}

As a concrete example to illustrate the strategy-grouping
phenomenon, we set $k = 5$. Figures \ref{fig:ts}(a-c) show time
series of $n_s$, the number of agents selecting each strategy $s$,
for $p=0$, 0.45, and 1.0, respectively. For Fig. \ref{fig:ts}(a)
where $p = 0$, an agent makes no informed decision in that it
changes strategy randomly with probability $m$ but stays with the
original strategy with probability $1-m$. In this case, $n_s$'s
appear random. For the opposite extreme case of $p = 1$ [Fig.
\ref{fig:ts}(c)], each agent makes well informed decisions based
on available local information about the strategies used by its
neighbors. In this case, the time series are quasiperiodic (a
detailed analysis will be provided in Sec. \ref{sec:theory}). For
the intermediate case of $p=0.45$ [Fig. \ref{fig:ts}(b)], agents'
decisions are partially informed. In this case, an examination of
the time series points to the occurrence of an interesting
grouping behavior: the 5 strategies, in terms of their selection
by the agents, are divided into two distinct groups $g_1$ and
$g_2$ that contain $k_{g_1}=2$ and $k_{g_2}=3$ resources,
respectively. The time series associated with the smaller group
exhibit larger fluctuations about its equilibrium.

\begin{figure}
\epsfig{figure=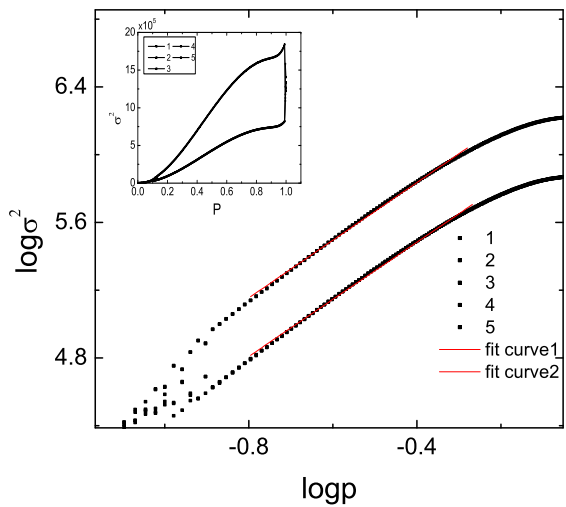,width=\linewidth}
\caption{For the same system in Fig. \ref{fig:ts}, variance $\sigma^2$
of the time series $n_s$ as a function of $p$ on a logarithmic scale.
A period-doubling like bifurcation occurs, at which a grouping
behavior emerges. Inset is the same plot but on a linear scale.}
\label{fig:variance}
\end{figure}

To better characterize the fluctuating behaviors in the time
series $n_s$, we calculate the variance
$\sigma^2_s=\langle[n_s(t)-N/k]^2\rangle$ as a function of the
system parameter $p$, where $\langle\cdot\rangle$ is the
expectation values averaged over a long time interval, as shown in
Fig. \ref{fig:variance} on a logarithmic scale. We observe a
generally increasing behavior in $\sigma^2_s$ with $p$ and,
strikingly, a bifurcation-like phenomenon. In particular, for $p <
p_b$, where $p_b$ is the bifurcation point, $\sigma^2$s for all
strategies assume approximately the same value. However, for $p >
p_b$, there are two distinct values for $\sigma^2$, signifying the
aforementioned grouping behavior [Fig. \ref{fig:ts}(b)]. From Fig.
\ref{fig:variance}, we also see that, after the bifurcation, the
two branches of $\sigma^2$ are linear (on a logarithmic scale) and
have approximately the same slope $a$, suggesting the following
power-law relation: $\sigma^2_{g_{i}}={b_{g_{i}}p^a}$, for $i =
1,2$, where $\log{(b_{g_1})}$ and $\log{(b_{g_2})}$ are the
intercepts of the two lines in Fig. \ref{fig:variance}. We thus
obtain
\begin{equation} \label{eq:main_ratio}
\frac{\sigma^2_{g_1}}{\sigma^2_{g_2}}=\frac{b_{g_1}}{b_{g_2}}.
\end{equation}
In Sec. \ref{sec:theory}, we will develop a theory to explain the
relations among the variances of the grouped strategies and to
provide formulas for the amplitudes
of the time series in Fig. \ref{fig:ts} and the sizes of the groups
(denoted by $k_{g_1}$ and $k_{g_2}$, respectively). Specifically, our theory predicts the
following ratio between the variances of the two bifurcated branches:
\begin{equation} \label{eq:main_ratio_theory}
\frac{\sigma^2_{g_1}}{\sigma^2_{g_2}}=(\frac{k_{g_2}}{k_{g_1}})^2,
\end{equation}
which is identical to the numerically observed ratio in
Eq. (\ref{eq:main_ratio}), with the additional prediction that the
strategies in the group of smaller size exhibit stronger fluctuations
since the corresponding value of $\sigma^2$ is larger. Overall, the
emergence of the grouping behavior in multiple-resource MGs, as
exemplified in Fig. \ref{fig:variance}, resembles a period-doubling
like bifurcation. While period-doubling bifurcations are extremely
common in nonlinear dynamical systems, to our knowledge, in complex
game systems a clear signature of such a bifurcation had not been
reported previously.

A careful examination of the time-series for various $p$ has revealed
that the strategy-grouping processes has already taken place prior to
the bifurcation point $p_b$ in the variance $\sigma^2$, but all
resulted grouping states are unstable. Take as an
example the $5$-strategy system in Fig. \ref{fig:ts}. In principle,
there can be two types of pairing groups: $(1,4)$ and $(2,3)$. For
any grouping state, the following constraint applies:
\begin{equation} \label{eq:constraint}
k_{g_1}+k_{g_2}=k, \qquad k_{g_1}\cdot k_{g_2}\neq 0
\end{equation}
There are in total $k/2$ (if $k$ is even) or $(k-1)/2$ (if $k$ is
odd) possible grouping states for the system with $k$ available
strategies. However, the grouping states are not stable for
$p<p_b$. What happens is that a strategy can remain in one group
but only for a finite amount of time before switching to a different
group. Assume that the sizes of the original two pairing groups
are $(k_{g_1}$ and $k_{g_2})$, respectively. The sizes of the new
pair of groups are thus $(k_{g_1}\pm1$ and $k_{g_2}\mp1)$, as
stipulated by Eq. (\ref{eq:constraint}). Associated with switching
to a different pair of groups, the amplitudes of the time series
$n_s$ for each strategy also change. As the bifurcation parameter
$p$ is increased, the stabilities of different pairs of grouping
states also change. At the bifurcation point $p_b$, one particular
pair of groups becomes stable, such as the grouping state $(2,3)$
in Fig. \ref{fig:variance}.

\begin{figure*}
\begin{center}
\epsfig{figure=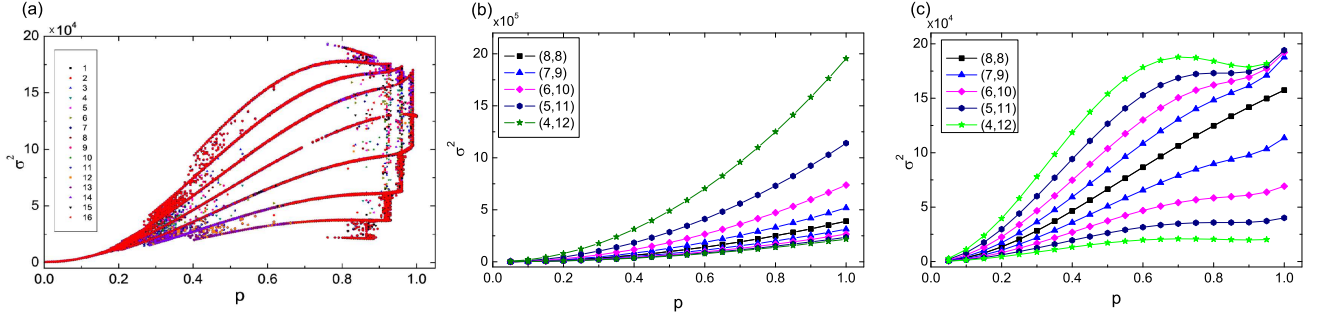,width=\linewidth}
\caption{(Color online.) For a multiple-resource MG system of
$N=10000$ agents on a square lattice and $k = 16$ available strategies,
(a) numerically obtained bifurcation-like behavior that leads
to the emergence of various pairs of grouping states [e.g, (8,8),
(4,12), etc.], (b) predicted bifurcation from mean-field theory
(Sec. \ref{sec:theory}), and (c) results from an improved mean-field
theory (Sec. \ref{sec:theory}). The probability of random selection
in the absence of local information is set to be $m = 1$.}
\label{fig:k16}
\end{center}
\end{figure*}

The bifurcation-like phenomenon and the emergence of various
strategy-grouping states are general for multiple-resource MG game
dynamics. For example, Fig. \ref{fig:k16} shows $\sigma^2$ as a
function of $p$ for a system with $k=16$ available strategies.
There are in total 8 possible grouping states, ranging from
$(8,8)$ to $(1,15)$. As $p$ is increased, the grouping states
$(8,8)$, $(7,9)$, $(6,10)$, $(5,11)$ and $(4,12)$ become stable
one after another, as can be seen from the appearance of their
corresponding branches in Fig. \ref{fig:k16}. The behavior can be
understood theoretically through a stability analysis (Sec.
\ref{sec:theory}).

Another phenomenon revealed by Fig. \ref{fig:k16} is the merging
of bifurcated branches. For example, as $p$ is increased through
about 0.8, the grouping states disappear one after another in the
reverse order as they initially appeared. This can also be
understood through the stability analysis (Sec. \ref{sec:theory}).

\section{Theory} \label{sec:theory}

Here we develop an analytic theory to understand the emergence,
characteristics, and evolutions of the strategy groups.

\subsection{Relationship among variance, amplitude and group size}

\begin{figure}
\epsfig{figure=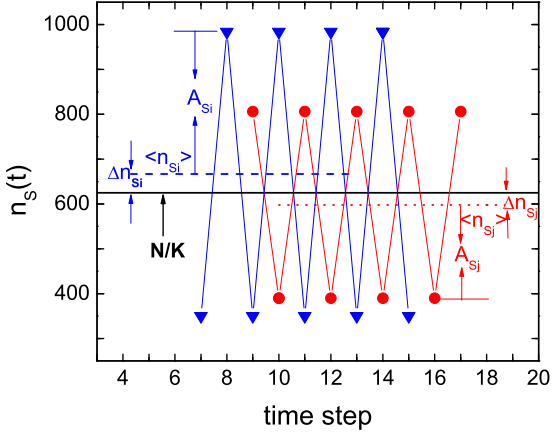,width=\linewidth}
\caption{For the case of two groups [e.g., Fig. \ref{fig:ts}(b)],
schematic illustration of time series $n_{s_i}$ in group
$g_1$ (blue circles) and $n_{s_j}$ in group $g_2$ (red triangles),
where various quantities such as the amplitude $A$ are labeled.}
\label{fig:sketchmap}
\end{figure}

In general, for a multiple-resource MG system of $N$ agents, as
the parameter $p$ is increased so that agents become more likely
to make informed decision for strategy selection, the available
strategies can be divided into pairs of groups. The example in
Fig. \ref{fig:ts}(b) presents a case where there are two distinct
strategy groups $g_1$ and $g_2$, which contain $k_{g_1}$ and
$k_{g_2}$ strategies, respectively, where $k_{g_1}+k_{g_2}=k$. For
Fig. \ref{fig:ts}(b), we have $k_{g_1}<k_{g_2}$. The strategies
belonging to the same group are selected by approximately the same
number of agents, i.e., the time series $n_s(t)$ for strategies in
the same group are nearly identical. During the time evolution, a
strategy $s$ can switch iteratively from being a minority strategy
[$n_s(t)<n_c \equiv N/k$] to being a majority one
[$n_s(t+1)>n_c$]. In particular, as shown in the schematic map in
Fig. \ref{fig:sketchmap}, for the strategy in group $g_1$ denoted
by $s_i$, and the strategy in group $g_2$ denoted by $s_j$, if
$n_{s_i}(t)<N/k<n_{s_j}(t)$, we will have
$n_{s_i}(t+1)>N/k>n_{s_j}(t+1)$. In addition, the time series
$n_s$ reveals that the average numbers of agents for strategies
$s_i$ and $s_j$, denoted by $\langle n_{s_i}\rangle$ and $\langle
n_{s_j}\rangle$ (the blue dash line and red dot line in Fig.
\ref{fig:sketchmap}), respectively, are not equal to $N/k$ (the
black solid line in Fig. \ref{fig:sketchmap}). In fact, we have
\begin{equation} \label{eq:I1}
\langle n_{s_i}\rangle=\frac{N}{k}+\Delta n_{s_i}, \quad \langle
n_{s_j}\rangle=\frac{N}{k}-\Delta n_{s_j}
\end{equation}
with $\langle n_{s_i}\rangle>N/k$, and $\langle n_{s_j}\rangle
<N/k$. Here, $\Delta n_{s_x}$ is the absolute value of the
difference between $\langle n_{s_x}\rangle$ and $N/k$. The number
of agents $n_{s_i}(t)$ [or $n_{s_j}(t)$] typically fluctuates about the
equilibrium $\langle n_{s_i}\rangle$ [or $\langle n_{s_j}\rangle$]
with amplitude $A_{s_i}$ [or $A_{s_j}$], as shown in the
schematic map in Fig. \ref{fig:sketchmap}.

Based on the numerical observations, we
can argue that the strategy grouping phenomenon is intimately related
to the fluctuations in the time series $n_s$. Assuming the MG system is
closed so that the number of agents is a constant, we have
\begin{equation} \label{eq:Conv}
k_{g_1}\cdot n_{s_i}(t)+k_{g_2}\cdot n_{s_j}(t)=N.
\end{equation}
Thus, for two consecutive time steps, we have,
\begin{eqnarray}
\nonumber
k_{g_1}\cdot (\langle n_{s_i}\rangle+A_{s_i})+k_{g_2}\cdot
(\langle n_{s_j}\rangle-A_{s_j}) & = & N, \\
k_{g_1}\cdot (\langle
n_{s_i}\rangle-A_{s_i})+k_{g_2}\cdot(\langle
n_{s_j}\rangle+A_{s_j}) & = & N. \label{eq:I2b}
\end{eqnarray}
Substituting Eq. (\ref{eq:I1}) into Eq. (\ref{eq:I2b}), we have
\begin{eqnarray}
\nonumber
k_{g_1}\cdot(\frac{N}{k}+\Delta n_{s_i}+A_{s_i})
& + & k_{g_2}\cdot
(\frac{N}{k}-\Delta n_{s_j}-A_{s_j}) =  N, \\ \nonumber
k_{g_1}\cdot(\frac{N}{k}+\Delta n_{s_i}-A_{s_i})
& + & k_{g_2}\cdot
(\frac{N}{k}-\Delta n_{s_j}+A_{s_j}) = N,
\end{eqnarray}
from which we obtain the relations between $A_{s_i}$ and
$A_{s_j}$, $\Delta n_{s_i}$ and $\Delta n_{s_j}$ as
\begin{equation} \label{eq:I4}
\frac{A_{s_i}}{A_{s_j}}=\frac {k_{g_2}}{k_{g_1}}, \quad
\frac{\Delta n_{s_i}}{\Delta n_{s_j}}=\frac {k_{g_2}}{k_{g_1}}.
\end{equation}
We see that the fluctuations of the time series are closely
related the grouping of the strategies.

From the definition of $\sigma^2_s$, we can write the variances of
$n_{s_i}$ and $n_{s_j}$ as
\begin{eqnarray}
\nonumber
\sigma_{s_i}^2 & = & \frac{(A_{s_i}+\Delta n_{s_i})^2+(A_{s_i}-\Delta
n_{s_i})^2}{2} = A_{s_i}^2+\Delta n_{s_i}^2, \\ \nonumber
\sigma_{s_j}^2 & = & \frac{(A_{s_j}-\Delta n_{s_j})^2+(A_{s_j}+\Delta
n_{s_i})^2}{2} = A_{s_j}^2+\Delta n_{s_j}^2.
\end{eqnarray}
Using Eq. (\ref{eq:I4}), we obtain
\begin{equation} \label{eq:I5}
\frac{\sigma_{s_i}^2}{\sigma_{s_j}^2}=(\frac{k_{g_2}}{k_{g_1}})^2.
\end{equation}
As shown in Fig. \ref{fig:variance}, the ratio of the variances of group
$g_1$ and $g_2$ from the simulation agree very well with Eq. (\ref{eq:I5}).

\subsection{Mean-field theory} \label{subsec:MF}

We develop a mean-field theory to understand the fluctuation patterns
of the system. To be concrete, we still treat the
case of two distinct groups. Consider strategy $s_i$ that belongs to
group $g_1$ and assume that $s_i$ is the majority strategy at time
$t=t_0$, i.e., $n_{s_i}^{(0)}>N/k$. According to the mean-field
approximation, at the next time step $t=t_0+1$, the number of
agents $n_{s_i}^{(1)}$ choosing strategy $s_i$ is
\begin{eqnarray} \label{eq:nsi}
n_{s_i}^{(1)} & = & n_{s_i}^{(0)}-(w_1+w_2)+w_3 \nonumber \\
& = & n_{s_i}^{(0)}-[n_{s_i}^{(0)}p + n_{s_i}^{(0)}(1-p)m\frac{k-1}{k}] \nonumber \\
& + & (N-n_{s_i}^{(0)})(1-p)m\frac{1}{k},
\end{eqnarray}
where
\begin{eqnarray}
\nonumber
w_1 & = & n_{s_i}^{(0)}p, \nonumber \\
w_2 & = & n_{s_i}^{(0)}(1-p)m\frac{(k-1)}{k}, \nonumber \\
w_3 & = & (N-n_{s_i}^{(0)})(1-p)m\frac{1}{k}.
\end{eqnarray}
Here, $w_1$ and $w_2$ together are the number of agents abandoning strategy $s_i$
(or the flow out of strategy $s_i$). That is, of the $n_{s_i}^{(0)}$ agents,
$w_1$ agents will act based on local information by selecting a minority
strategy different than $s_i$ for the next time step $t_0+1$. At the same
time, there will be $w_2$ agents acting without local information by choosing
randomly one of the other $k-1$ strategies. The quantity $w_3$ represents the
flow into $s_i$ from the remaining $N-n_{s_i}^{(0)}$ agents.
These agents will mutate randomly to
switch their strategies to $s_i$ without any local information. We thus have
\begin{equation} \label{eq:nsi+}
n_{s_i}^{(1)}=n_{s_i}^{(0)}(1-p)(1-m)+N(1-p)m\frac{1}{k}.
\end{equation}
Suppose $n_{s_i}^{(1)}<N/k$. Then $s_i$ and the other $k_{g_1}-1$ strategies
in $g_1$ are the minority strategy, and the agents selecting those strategies
win the game at $t=t_0+1$. The time series $n_{s_i}^{(2)}$ at time $t_0+2$
can be written as
\begin{eqnarray}
n_{s_i}^{(2)}& = & n_{s_i}^{(1)}-n_{s_i}^{(1)}[p\frac{(k_{g_1}-1)}
{k_{g_1}}+(1-p)m\frac{(k-1)}{k}] \nonumber \\
& + & (N-n_{s_i}^{(1)})[p\frac{1}{k_{g_1}}+(1-p)m\frac{1}{k}] \nonumber\\
& = &n_{s_i}^{(1)}-(w_4+w_5)+(w_6+w_7), \label{eq:nsi2}
\end{eqnarray}
where $w_4$ and $w_5$ stand for the flows out of strategy $s_i$, while
$w_6$ and $w_7$ represent the flows into $s_i$ from those
$N-n_{s_i}^{(0)}$ agents on other strategies. We have
\begin{equation} \label{eq:nsi++}
n_{s_i}^{(2)}=n_{s_i}^{(1)}(1-p)(1-m)+N(1-p)m\frac{1}{k}+Np\frac{1}{k_{g_1}},
\end{equation}
where $n_{s_i}^{(2)}$ is larger than $N/k$, $s_i$ and all other strategies
in group $g_1$ will be the majority strategy again, as at time $t_0$.
The process $n_{s_i}\rightarrow n_{s_i}^{(1)} \rightarrow n_{s_i}^{(2)}$
thus occurs iteratively. From Eqs. (\ref{eq:nsi+}) and (\ref{eq:nsi++}),
we can get the number of agents for one given strategy at any time $t$.
In particular, denoting $\alpha \equiv (1-p)(1-m)$, $\beta \equiv N(1-p)m/k$,
and $\gamma_{g_1} \equiv Np/k_{g_1}$, we obtain the iterative dynamics for
agents selecting strategy $s_i$ as
\begin{eqnarray} \label{eq:nsiB}
&&n_{s_i}^{(1)}=\alpha n_{s_i}^{(0)}+\beta, \\ \nonumber
&&n_{s_i}^{(2)}=\alpha n_{s_i}^{(1)}+\beta+\gamma_{g_1}, \\ \nonumber
&&n_{s_i}^{(3)}=\alpha n_{s_i}^{(2)}+\beta, \\ \nonumber
&&\cdots \cdots \cdots \cdots \cdots \\ \nonumber
&&n_{s_i}^{(2a)}=\alpha n_{s_i}^{(2a-1)}+\beta+\gamma_{g_1}, \\ \nonumber
&&n_{s_i}^{(2a+1)}=\alpha n_{s_i}^{(2a)}+\beta, \\ \nonumber
&&\cdots \cdots \cdots \cdots \cdots
\end{eqnarray}
where $a\in R$ and $n_{s_i}^{(t')}$ stands for the number of
agents at $t=t_0+t'$. Carrying out the iterative process in
Eq. (\ref{eq:nsiB}), we obtain
\begin{eqnarray}
\nonumber
&&n_{s_i}^{(2)}=\alpha^2 n_{s_i}^{(0)} + \alpha\beta+\beta+\gamma_{g_1}, \\ \nonumber
&&n_{s_i}^{(3)}=\alpha^2 n_{s_i}^{(1)} + \alpha\beta+\alpha \gamma_{g_1}+\beta.
\end{eqnarray}
For the case where $n_s(t)$ exhibits stable oscillations, i.e., the system is
in a {\em stationary state}, we have,
\begin{eqnarray}
\nonumber
&&n_{s_i}^{(0)}=n_{s_i}^{(2)}=\cdots=n_{s_i}^{(2a)} \quad a\in R, \\ \nonumber
&&n_{s_i}^{(1)}=n_{s_i}^{(3)}=\cdots=n_{s_i}^{(2a+1)}.
\end{eqnarray}
We thus obtain the values of $n_{s_i}^{(2a)}$ and
$n_{s_i}^{(2a+1)}$ as a function of the probability $p$, mutation
probability $m$, and the grouping parameter $k_{g_1}$, and $k$:
\begin{eqnarray}
\nonumber
&&n_{s_i}^{(2a)}=\frac{\alpha\beta+\beta+\gamma_{g_1}}{1-\alpha^2}, \\
&&n_{s_i}^{(2a+1)}=\frac{\alpha\beta+\alpha
\gamma_{g_1}+\beta}{1-\alpha^2}. \label{eq:nsi2a+1}
\end{eqnarray}
From $n_s$, we can get the expression of
the amplitude of the fluctuation, the mean value
$\langle{n_{s_i}}\rangle$, and its difference from $N/k$ as,
\begin{eqnarray}
\nonumber
&&A_{s_i}=[n_{s_i}^{(2a)}-n_{s_i}^{(2a+1)}]/2=\frac{\gamma_{g_1}}{2(1+\alpha)}, \\ \nonumber
&&\langle{n_{s_i}}\rangle=[n_{s_i}^{(2a)}+n_{s_i}^{(2a+1)}]/2
=\frac{2\beta+\gamma_{g_1}}{2(1-\alpha)}, \\ \nonumber
&&\Delta
n_{s_i}=|\langle{n_{s_i}}\rangle-N/k|=|\frac{k\gamma_{g_1}-2Np}{2k(1-\alpha)}|
=\frac{Np|k/k_{g_1}-2|}{2k(1-\alpha)}.
\end{eqnarray}
In the above derivation, we have assumed that $s_i$ belongs to group $g_1$,
and obtained expressions of the $n_{s_i}$, $A_{s_i}$, $\langle{n_{s_i}}\rangle$,
and $\Delta n_{s_i}$. Similarly, using Eq. (\ref{eq:Conv}), we can calculate the time
series $n_{s_j}$, the number of agents selecting the strategy $s_j$ belonging
to group $g_2$, and the corresponding characterizing quantities.

Alternatively, following the steps similar to those from Eqs. (\ref{eq:nsi})
to (\ref{eq:nsiB}), we can write the recurrence formula for $n_{s_j}$ as
\begin{eqnarray}
\nonumber
&&n_{s_j}^{(2a)}=\alpha n_{s_j}^{(2a-1)}+\beta, \\ \nonumber
&&n_{s_j}^{(2a+1)}=\alpha n_{s_j}^{(2a)}+\beta+\gamma_{g_2},
\end{eqnarray}
where $\gamma_{g_2}=Np/k_{g_2}$. For the case of stable strategy, we
get the corresponding number of agents as
\begin{eqnarray}
\nonumber
&&n_{s_j}^{(2a)}=\frac{\alpha\beta+\alpha
\gamma_{g_2}+\beta}{1-\alpha^2}, \\
&&n_{s_j}^{(2a+1)}=\frac{\alpha\beta+\beta+\gamma_{g_2}}{1-\alpha^2}.
\label{eq:nsj2a+1}
\end{eqnarray}
We find that the values of $n_s$ obtained from Eq. (\ref{eq:nsj2a+1}) agree with
those from Eq. (\ref{eq:Conv}) very well (derivation and data not shown).
In addition, the expressions of $A_{s_j}$, $\langle{n_{s_j}}\rangle$, and
$\Delta n_{s_j}$ are identical to those associated with $s_i$, with
the quantity $\gamma_{g_1}$ replaced by $\gamma_{g_2}$.

\begin{figure}
\epsfig{figure=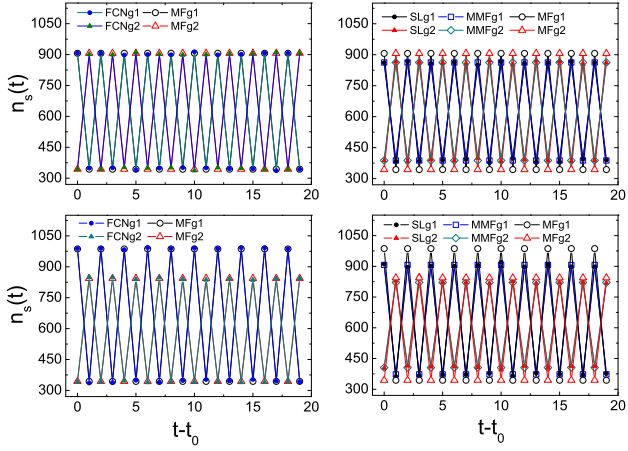,width=\linewidth}
\caption{ (Color online.) Comparison of the time series of
$n_{s_i}$ in group $g_1$ (and $n_{s_j}$ in group $g_2$) from the
mean-field formula [Eq. (\ref{eq:nsi2a+1})] and modified
mean-field formula [Eq. (\ref{eq:MFn})], and from simulations. The
upper and lower figures show the results for the grouping state
$(8,8)$  and $(7,9)$, respectively. The left panels are the
results from mean-field theory (MF) and from simulations on fully
connected networks (FCN). The right panels are the results from
our modified mean-field theory (MMF) and simulations on the square
lattice (SL). The system parameters are $N=10000$, $k=16$,
$p=0.45$, and $m=1$.} \label{fig:MF}
\end{figure}

The mean-field theory is ideally suited for fully connected
networks. Indeed, results from the theory and direct simulations
agree with each other very well, as shown in Fig. \ref{fig:MF}.
However, in real-world situations, a fully connected topology
cannot be expected, and the mean-field treatment will no longer
be accurate. For example, we have carried out simulations on
square-lattice systems and found noticeable deviations from the
mean-field prediction. To remedy this deficiency, we develop
a modified mean-field analysis for MG dynamics on sparsely
homogeneous networks (e.g., square lattices or random networks).

\subsection{Mean-field theory for MG dynamics on sparsely
homogeneous networks} \label{subsec:MMF}

Due to the limited number of links in a typical large-scale network, it
is possible for a failed agent to be surrounded by agents
from the same group (who will likewise fail the game). In this case,
the failed agent has no minority strategy to imitate (set
$\Pi$ is empty) and thus will randomly select one strategy from
the $k$ available strategies. Taking this effect into account,
we can modify the mean-field approximation in Eqs. (\ref{eq:nsi})
and (\ref{eq:nsi2}) as
\begin{eqnarray} \label{eq:MF22}
&&n_{s_i}^{(1)}=n_{s_i}^{(0)}-(w_1+w_2)+w_3+w_8, \\ \nonumber
&&n_{s_i}^{(2)}=n_{s_i}^{(1)}-(w_4+w_5)+(w_6+w_7)+w_9,
\end{eqnarray}
with the two modified terms given by
\begin{eqnarray}
\nonumber
&&w_8=k_{g_1}n_{s_i}^{(0)}p\eta_{g_1}\frac{1}{k}, \\ \nonumber
&&w_9=(N-k_{g_1}n_{s_i}^{(1)})p\eta_{g_2}(\frac{1}{k}-\frac{1}{k_{g_1}}),
\end{eqnarray}
where $\eta_{g_x}$ is the probability for one agent in group $g_x$
to be surrounded by agents from the same group. The quantity $w_8$ stands
for the flow from the failed agents in group $g_{1}$ [the number
is $N_{g_1}\equiv k_{g_1}n_{s_i}^{(0)}$], who react to the
information [the number is $N_{g_1}p$] but with no winner
surrounded to supply the optional minority strategy [the number is
$N_{g_1}p\eta_g$], and thus select $s_i$ with probability $1/k$.
The quantity $w_9$ represent two factors: (1) the
$n_1=(N-k_{g_1}n_{s_i}^{(1)})p\frac{1}{k_{g_1}}\eta_{g_2}$ failed agents
in group $g_2$ who should have flow into $s_i$ [i.e. $w_6$] but
are held back because they are surrounded by agents in $g_2$, and
(2) the failed agents in group $g_2$ who are surrounded by agents
in $g_2$ and thus select $s_i$ with $1/k$ probability, the
number of which is $n_2=(N-k_{g_1}n_{s_i}^{(1)})p\eta_{g_2}\frac{1}{k}$.
Apparently, we have $w_9=n_2-n_1$.

From Eq. (\ref{eq:MF22}), we obtain
\begin{eqnarray} \label{eq:MFn}
&&n^{(2a)}_{s_i}=\frac{\alpha''_{g_1}\beta+\beta+\gamma'_{g_1}}
{1-\alpha'_{g_1}\alpha''_{g_1}}, \\ \nonumber
&&n^{(2a+1)}_{s_i}=\frac{\alpha'_{g_1}\beta+\alpha'_{g_1}\gamma'_{g_1}+\beta}
{1-\alpha'_{g_1}\alpha''_{g_1}}, \\ \nonumber
&&A_{s_i}=\frac{(\alpha''_{g_1}-\alpha'_{g_1})\beta+(1-\alpha'_{g_1})\gamma'_{g_1}}
{2(1+\alpha'_{g_1}\alpha'')}, \\ \nonumber
&&\langle{n_{s_i}}\rangle=\frac{\beta(\alpha''_{g_1}+\alpha'_{g_1}+2)+\gamma'_{g_1}
(1+\alpha'_{g_1})}{2(1-\alpha'_{g_1}\alpha''_{g_1})}, \\ \nonumber
&&\Delta
n_{s_i}=|\frac{\beta(\alpha'_{g_1}+\alpha''_{g_1})+\gamma'_{g_1}(1+\alpha'_{g_1})
+2\beta}{2(1-\alpha'_{g_1}\alpha''_{g_1})}-\frac{N}{k}|,
\end{eqnarray}
where the parameters are
\begin{eqnarray}
&&\alpha'_{g_1}=(1-p)(1-m)+\frac{k_{g_1}p\eta_{g_1}}{k}, \nonumber\\
&&\alpha''_{g_1}=(1-p)(1-m)+k_{g_1}p\eta_{g_2}(\frac{1}k_{g_1}-\frac{1}{k}),\nonumber\\
&&\gamma'_{g_1}=Np[\frac{1}{k_{g_1}}+\eta_{g_2}(\frac{1}{k}-\frac{1}{k_{g_1}})]. \nonumber
\end{eqnarray}
The equation set (\ref{eq:MFn}) represents the modified mean-field
description of the time series associated with the stable
strategies in the game system supported on sparsely homogeneous
networks. The density of agents in $g_x$ is denoted by
$\rho_{g_{x}}\equiv N_{g_{x}}/N$. For the case where agents from
different groups are well mixed in the network, the probability
for one given agent in $g_x$ to meet with agents in $g_y$ is
$P_{g_{x}g_{y}}=\rho_{g_{y}}$ (for $x, y= 1, 2$). If the average
degree of the network is $d$, the probability that one agent from
$g_x$ is surrounded by agents from the same group is
$\eta_{g_x}=(P_{g_{x}g_{x}})^d=\rho_{g_{x}}^d$. From the
simulation on the square lattice system where each agent has $d=4$
neighbors, we observe a quite weak effect of clustering of agents
in $g_1$ or $g_2$, so $\eta_{g_x}\approx
\rho_{g_{x}}^4=(N_{g_{x}}/N)^4$. Based on the quantity
$\eta_{g_x}$, analytic prediction of the time series in the
modified mean-field (MMF) theory can be obtained, as shown in Fig.
\ref{fig:MF} for a square-lattice system. We observe a good
agreement with simulation results. Simulations on homogeneous
networks of different values $d$ have also been carried out, with
results in good agreement with the prediction from the modified
mean-field theory.

\subsection{Stability of strategy grouping states}

\begin{figure}
\epsfig{figure=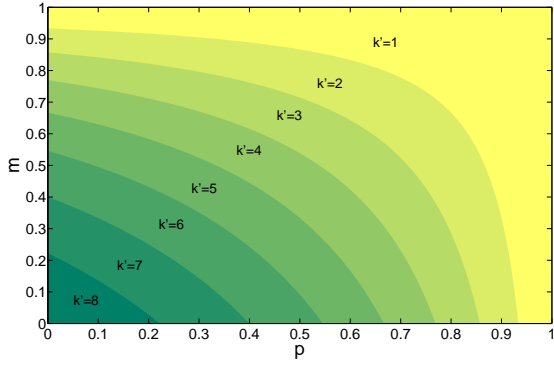,width=\linewidth}
\caption{(Color online.) Phase diagram indicating the stability region
of strategy-grouping states. The stable parameter region of one given
grouping state with $k_{g_1}=k'$ is at the upper right of the
corresponding curve of $k'$.}
\label{fig:pm}
\end{figure}

Our mean-field treatment yields formula characterizing the stable
oscillations associated with the grouping state $(k_{g_1},
k_{g_2})$, which include the variance ratio of the pairing groups
[Eq. (\ref{eq:I5})] and time series $n_{s}$ [Eqs.
(\ref{eq:nsi2a+1}) and (\ref{eq:MFn})]. However, from the
simulation result shown in Fig. \ref{fig:k16}, we see that not all
the grouping states are stable in the parameter space. As $p$ is
increased, the strategy-grouping state of the smaller $k_{g_1}$
becomes stable, and the corresponding branch appears. It is
therefore useful to analyze the stability of the grouping state.

In our treatment we have assumed $n_{s_i}^{(0)}>N/k$ and
$n_{s_j}^{(0)}<N/k$. Then, the necessary condition for the
grouping state $(k_{g_1}, k_{g_2})$ to become stable is
$n_{s_i}^{(1)}<N/k$ and $n_{s_j}^{(1)}>N/k$. Using Eqs.
(\ref{eq:nsi2a+1}) and (\ref{eq:nsj2a+1}), we get
\begin{eqnarray} \label{eq:stability_1}
k_{g_1} & > & \frac{k(1-m-p+pm)}{2-m-p+pm}\equiv\xi_1(p,m), \\ \nonumber
k_{g_2} & < & \frac{k}{2-m-p+mp}\equiv\xi_2(p,m),
\end{eqnarray}
where $\xi_1$ and $\xi_2$ are continuous functions of the
parameters $p$ and $m$, and $\xi_1+\xi_2=k$. The two inequalities
in Eq. (\ref{eq:stability_1}) are nevertheless equivalent to each
other. Figure \ref{fig:pm} presents a phase diagram in the
parameter space, where the curves of $\xi_1(p,m)=k'$ for $k'=8$,
$7$, $\cdots$, $1$ are shown. The necessary condition for the
strategy-grouping state with $k_{g_1}=k'$ to be stable is that the
the parameters $p$ and $m$ are in the upper-right region of the
curve $\xi_1(p,m)=k'$. For certain value of $m$, only when
$p>p_b(m,k')$ will the state of $k_{g_1}=k'$ be stable. While the
value of $p_b$ from simulation is different from the theoretical
value $p_b(m,k')$, our mean-field theory does provide a
qualitative explanation for the phenomenon in Fig.
\ref{fig:k16}(a), where more branches of smaller strategy-grouping
states become stable as $p$ is increased.

We have also seen in Fig. \ref{fig:variance}, and Fig.
\ref{fig:k16}(a) that, as $p$ approaches $1$, the bifurcated
branches of different grouping state merge together. For the case
of even $k$, $n_s$ fluctuates stably in the grouping state with
$k_{g_{1}}=k_{g_{2}}=k/2$. While, for the case of odd $k$, $n_s$
fluctuates quasiperiodicly [see Fig. \ref{fig:ts}(c)]. Actually,
the grouping state always switches between $(k_{g}+1, k_{g})$ and
$(k_{g}, k_{g}+1)$, with $k_{g}=(k-1)/2$. We can understand the
instability and merging of grouping states from Fig.
\ref{fig:ts}(c), and the schematic map in Fig.
\ref{fig:sketchmap}, as follows. As $p$ is increased to $1$,
$\Delta n_{s_i}$ and $\Delta n_{s_j}$ increase and become
comparable to the amplitudes $A_{s_i}$ and $A_{s_j}$,
respectively. Namely, the attendances $n_s$ of strategies can be
very close to $n_c=N/k$ . In case that $n_s$ of one strategy does
not get cross $n_c$ because of \emph{noise}, i.e., it acts as
minority (or majority) strategy twice, then, the fluctuation of
$n_s$, as well as the grouping state is changed.

\begin{figure*}
\epsfig{figure=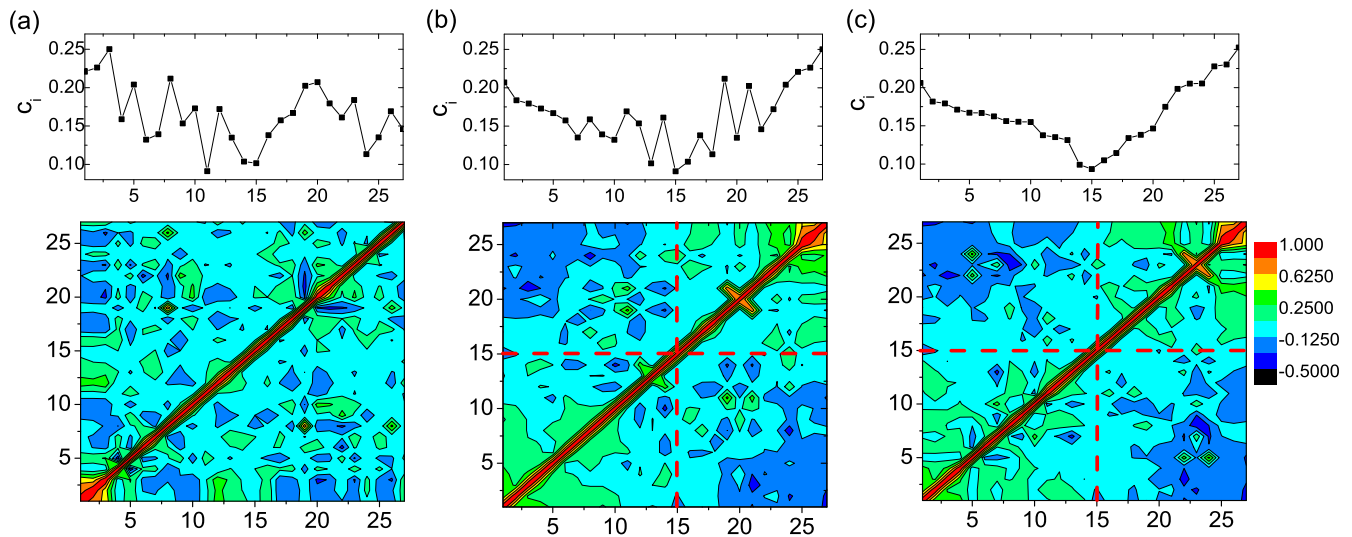,width=\linewidth}
\caption{(Color online.) Correlation matrix of log-returns for 27
stocks in Shanghai Stock Market's Steel Plate. (a) Original matrix
$\textbf{C}$, (b) matrix $\textbf{C}'$ ordered according to the
eigenvector for the maximum eigenvalue of $\textbf{C}$, and (c)
reordered matrix $\textbf{C}''$ from $\textbf{C}'$ with respect to
$c_i$ within each group. The $c_i$ values of the stocks are shown
for each matrix.} \label{fig:matrix}
\end{figure*}

\section{A real-world example: emergence of grouping states in
financial market} \label{sec:real_world_example}

The financial market is a representative multi-resource complex
system, in which many stocks are available for investment. We
analyze the fluctuation of the stock price from the empirical data
of 27 stocks in the Shanghai Stock Market's Steel Plate between
2007 and 2010. We regard the $27$ stocks, which are derived from
the iron and steel industry, as constituting a MG system with $k =
27$ resources, where the agents selecting the resources correspond
to the capitals invested. This system is open in the sense that
capital typically flows in and out, which is the main difference
from our closed-system model. In particular, given the time series
$x_i(t)$ of the daily closing price of stock $i$, the daily
log-return is defined as $R_i(t)=\ln x_i(t)-\ln x_i(t-1)$. The
average return of the $27$ stocks at time $t$, denoted by $\langle
R_i(t)\rangle$, signifies a global trend of the system at $t$,
which is caused by the change in the total mount of the capital in
this open, 27-stock system. However, when we analyze the detrended
log-returns $R_i'(t)=R_i(t)-\langle R_i(t)\rangle$, the system
resembles a closed system, as our model MG system. We shall
demonstrate that the strategy-grouping phenomenon occurs in this
real-world system.

We calculate the Pearson parameter $c_{ij}$ of each pair of the
detrended log-returns $R_i'(t)$ and $R_j'(t)$, which leads to a
$k\times k$ correlation matrix $\textbf{C}$, as shown in Fig.
\ref{fig:matrix}(a). In terms of the eigenvector associated with
the maximum eigenvalue of matrix $\textbf{C}$, we rank the order
of the stocks and obtain the matrix $\textbf{C}'$, as shown in
Fig. \ref{fig:matrix}(b). {\em The striking behavior is that the
matrix is apparently divided into 4 blocks, a manifestation of the
grouping phenomenon.} In particular, the matrix elements $c_{ij}$
among the first 15 stocks and those among the remaining 12 stocks
are generally positive, but the cross elements between stocks in
the two groups are negative. It is thus quite natural to classify
the first 15 stocks as belonging to group $g_1$ and the remaining
12 to group $g_2$. We can then write the matrix in a block form as
\begin{equation}
\nonumber
\textbf{C}'=\begin{array}{ccc}
\left(\begin{array}{cc}\textbf{C}_{g_2g_1}&\textbf{C}_{g_2g_2} \\
\textbf{C}_{g_1g_1}&\textbf{C}_{g_1g_2}\end{array}\right),
\end{array}
\end{equation}
where the elements of $\textbf{C}_{g_1g_1}$
and $\textbf{C}_{g_2g_2}$ are positive, and those of
$\textbf{C}_{g_1g_2}$ and $\textbf{C}_{g_2g_1}$ are negative.
The phenomenon is that the $27$-stock system has self-organized
itself into a $(12,15)$ grouping state, which is a natural consequence
of the MG dynamics in multi-resource complex systems.

For one given stock $i$, the mean absolute correlation is
\begin{equation}
\nonumber
c_i=\sum_{j=1}^{k}{|c_{ij}|}/(k-1), \qquad j\neq i.
\end{equation}
This parameter reflects the weight of the stock in the system. If
$c_i\rightarrow 0$, oscillations of stock $i$ are contained in the
noise floor. In this case, there is no indication as to whether
this stock belongs to group $g_1$ or $g_2$. The larger the value
of $c_i$, the less ambiguous that the stock belongs to either one
of the two groups. From the value of $c_i$ ranked in the same order
as in $\textbf{C}'$, we can see that the boundary of the
two groups is the stock with minimum $c_i$. Thus $c_i$ can be
considered as the characteristic number to distinguish different
groups. We can also reorder the matrix $\textbf{C}'$ according to
$c_i$ within group $g_1$ and $g_2$, respectively. This leads to
the matrix $\textbf{C}''$, as shown in Fig. \ref{fig:matrix}(c),
further demonstrating the grouping phenomenon.


\begin{figure*}
\epsfig{figure=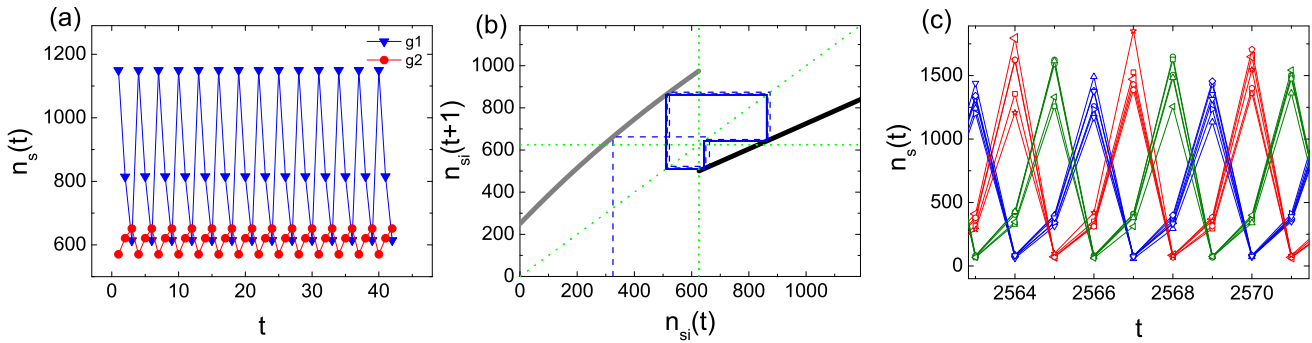,width=\linewidth}
\caption{(Color online.) (a) The analytical time series $n_s(t)$
of period-tripling in the form of $2$ groups, and (b) the
corresponding mapping from $n_{s_i}(t)$ to $n_{s_i}(t+1)$ for the
strategy $s_i$ in group $g_1$ [see Eq. (\ref{eq:MF22})]. The
system is of $k=16$, $p=0.25$, $m=0.2$, and the grouping state
$(k_{g_1}, k_{g_2})$ is $(3,13)$, respectively. (c) The time
series $n_s(t)$ of the stable grouping state with period-tripling
in the form of $3$ groups from simulation on FCN. The system is of
$k=16$, $p=0.88$, $m=1$, and the grouping state $(k_{g_1},
k_{g_2}, k_{g_3})$ is $(5,5,6)$.} \label{fig:mapping}
\end{figure*}

\section{Conclusions and discussions} \label{sec:conclusion}

Minority game, since its invention about two decades ago, has
become a paradigm to study the social and economical phenomena
where a large number of agents attempt to make simultaneous
decision by choosing one of the available options
\cite{Gintis:2009}. In the most commonly studied case of a single
available resource with players' two possible options, agents
taking the minority option are the guaranteed winners. Various
minority game dynamics have also received attention from the
physics community due to their high relevance to a number of
phenomena in statistical physics. It has become more and more
common in the modern world that multiple resources are available
for various social and economical systems. If the rule still holds
that the winning options are minority ones, the questions that
naturally arise are what type of collective behaviors can emerge
and how they would evolve in the underlying complex system. Our
present work aims to address these questions computationally and
analytically.

The main contribution and findings of this paper are the
following. Firstly, we have constructed a class of spatially
extended systems in which any agent interacts with a finite but
fixed number of neighbors and can choose either to follow the
minority strategy based on information about the neighboring
states or to select one randomly from a set of available
strategies. The probability to follow the local minority strategy,
or the probability of minority preference, is a key parameter
determining the dynamics of the underlying complex system.
Secondly, we have carried out extensive numerical simulations and
discovered the emergence of a striking collective behavior: as the
minority-preference probability is increased through a critical
value, the set of available strategies/resources spontaneously
break into pairs of groups, where the strategies in the same group
are associated with a specific fluctuating behavior of attendance.
This phenomenon of strategy-grouping is completely self-organized,
which we conjecture is the hallmark of MG dynamics with multiple
resources. Thirdly, we have developed a mean-field theory to
explain and predict the emergence and evolution of the
strategy-grouping states, with good agreement with the numerics.
Fourthly, we have examined a real-world system of a relatively
small-scale stock-trading system, and found unequivocal evidence
of the grouping phenomenon. Our results suggest grouping of
resources as a fundamental type of collective dynamics in
multiple-resource MG systems. Other real-world systems for which
our model is applicable include, e.g., hedge-fund portfolios in
financial systems, routing issues in computer networks and urban
traffic systems. We expect our model and findings to be not only
relevant with statistical-physical systems, but also important to
a host of social, economical, and political systems.

Additionally, the double-grouping (or paired grouping)
period-doubling like bifurcation are not the exclusive mode for
the grouping of resources. We have also observed the period$-3$
double-grouping [see Fig. \ref{fig:mapping}(a)(b)], and the
period$-3$ triplet-grouping bifurcation phenomena [see Fig.
\ref{fig:mapping}(c)]. Further research about these phenomena are
of great interest and is valuable for the understanding of chaotic
and bifurcation in social systems.

\section{Acknowledgement}

ZGH thanks Prof. Matteo Marsili for helpful discussions. This work
was partially supported by the NSF of China (Grant Nos. 10905026,
10905027, 11005053, and 11135001), SRFDP No. 20090211120030, FRFCU
No. lzujbky-2010-73, Lanzhou DSTP No. 2010-1-129. YCL was
supported by AFOSR under Grant No. FA9550-10-1-0083.

\end{document}